\documentclass[12pt]{iopart}
\usepackage{cite,graphicx}
\usepackage{amssymb}
\usepackage{iopams}
\usepackage{setstack}

\begin{document}

\title{Quantum suppression of chaotic tunneling}
\author{Akiyuki Ishikawa, Atushi Tanaka and Akira Shudo}
\address{Department of Physics, Tokyo Metropolitan University,
	Minami-Osawa, Hachioji, Tokyo 192-0397, Japan}

\begin{abstract}

The interplay between chaotic tunneling and dynamical localization 
in mixed phase space is investigated.  Semiclassical analysis 
using complex classical orbits reveals that tunneling through torus regions 
and transport in chaotic regions are not independent processes, 
rather they are strongly correlated and 
described by complex orbits with both properties.  
This predicts a phenomenon analogous to the quantum suppression of
classical diffusion: chaotic tunneling is suppressed as a result of 
dynamical localization in chaotic regions.  This hypothesis is 
confirmed by numerical experiments where the effect of destructive 
interference is attenuated.

\end{abstract}


Tunneling phenomena are purely quantum effects. Nevertheless its nature is strongly 
influenced by underlying classical dynamics.  In particular nontrivial aspects appear 
in the dynamical tunneling in mixed-type phase space, 
in which quasi-periodic and chaotic trajectories coexist \cite{BTU}.  
Tunneling transitions between quasi-doublet states are enhanced 
by chaotic states \cite{TU} and the existence of nonlinear resonances 
also leads to a qualitative 
change of tunneling processes 
\cite{BSU,mouchet}.  
The problem of quantum tunneling in multidimensional 
systems or more specifically in nonintegrable systems 
first raised in \cite{W} would be of 
fundamental importance in quantum mechanics,
and an approach taken there has recently been extended in 
\cite{Creagh0}. 
However, our understanding for multidimensional tunneling 
is still far from complete.  
A primary difficulty lies in the fact that dynamical tunneling 
in chaotic systems takes place in very complicated phase space; 
the structure of classical phase space itself is not 
an easily understandable object.  Another reason would be that dynamical tunneling proceeds  
in complex environments in the sense of wave phenomena.  Scarring \cite{Heller} 
or dynamical localization \cite{Casati,FGP}, 
or other types of invariant structure become sources 
of partial structures often observed in wave functions. 
It is not clear at all to what extent these various wave effects 
are independent of each other. 
These aspects make it difficult to evaluate the tunneling rate between 
torus to chaotic regions quantitatively. 

The trajectory description would be one of promising strategies to understand quantum phenomena 
of chaotic systems. In particular, the semiclassical analysis is now recognized as 
an efficient approach to this end, and there are indeed 
a bunch of numerical tests supporting its validity.  Concerning dynamical tunneling, it was shown 
that the complex semiclassical theory works fairly well and explains the mechanism of tunneling 
penetration out of quasi-periodic regions to chaotic seas \cite{Creagh,SI,TI}. 
In such a treatment, the classical dynamics is extended to complex phase space, and the trajectories 
on the Julia set is most responsible for reproducing tunneling wavefunctions 
 \cite{SII}. 

The aim of the present Letter is, based on the arguments predicted by the complex trajectory description of chaotic tunneling, to show that dynamical tunneling in mixed phase space and dynamical 
localization are strongly correlated to each other, 
so that the destruction of coherence, 
or more precisely the destruction of destructive interference in chaotic regions,  
not only induces delocalization of wavefunction in chaotic regions, 
but also causes strong enhancement of tunneling transition.  
In other words, $\lq\lq$genuine" chaotic tunneling is suppressed by dynamical 
localization in surrounding chaotic regions, which is entirely analogous mechanism 
as dynamical localization suppresses classical diffusion \cite{Casati,FGP}. 
The result must serve as further understandings of {\it amphibious states} \cite{Ketz1,Ketz2}, 
recently discovered quantum states that show the failure of 
semiclassical wavefunction hypothesis \cite{Percival}.

\begin{figure}
\begin{center}
\includegraphics[width=.70\linewidth]{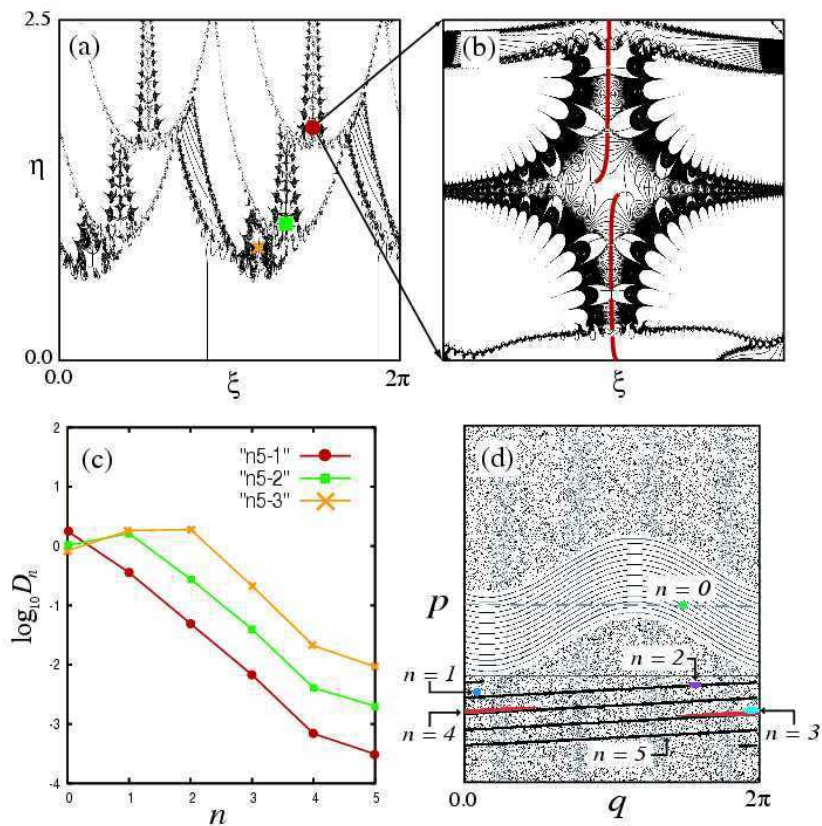}
\end{center}
\caption{
(a) A set of initial conditions ${\cal M}_n^{\alpha}$ ($\alpha = 0$), and 
(b) its magnification.  Red lines are parts of chained structures which 
substantially contribute to the semiclassical propagator  (\ref{semicl}). 
$|{\rm Im}q_n|$ of such orbits are very small. 
The parameters are chosen as $n = 5, a = 5, b=100$, 
$d_1 = -24, d_2 = -26, \omega = 1, p_d = 5$ and $K = 2$. 
(c) The distance from the real plane ($D_n = \sqrt{|p_n|^2+|q_n|^2}$) as a 
function of time step, where $p_n$ and $q_n$ denote $(p,q)$ at $n$. 
The initial conditions are taken from the points indicated in (a). 
Note that the orbits whose initial imaginary parts are small (yellow and green) are not 
necessarily go to the real plane directly, rather take a side trip in complex phase 
space. 
(d) The Lagrangian manifold projected onto real phase space. Here the Lagrangian manifold refers to $F^n(M_n^{\alpha})$, and the initial conditions giving these manifolds shown in (d) are red parts shown in (b).  The green dot (labeled as $n=0$) represents 
the initial condition and black curves (labeled as $n=5$) is the final Lagrangian manifold. 
 }
\label{fig1}
\end{figure}

We first provide evidence for why we can predict that chaotic tunneling is tightly 
correlated with the dynamical localization process.  
As found in \cite{SI,SII}, an exponentially large number of complex orbits appear as 
the contributors of the time-domain semiclassical propagator, and they indeed have almost 
equal weights in the semiclassical sum, which means that chaotic tunneling occurs as 
a consequence of superposition of exponentially many component waves. 
To demonstrate it,  we here employ an area-preserving map
\begin{eqnarray}\label{map}
\displaystyle
F: \left(\begin{array}{c}p' \\ q'\end{array}\right)
=
\left(\begin{array}{c}p - V'(q) \\ q + T'(p')\end{array}\right) 
\end{eqnarray}
where 
\begin{eqnarray} \nonumber
T'(p) &=& a p +  \frac12 (d_1 - d_2) 
+\frac{1}{2} \bigl[ a p - \omega + d_1 \bigr]
\tanh b (p-p_d)  \\ \label{kinetic}
&+& \frac{1}{2} \bigl[ -a p + \omega + d_2  \bigr] \tanh b (p+p_d) \\
\label{potential}
V'(q) &=& -K \sin q . 
\end{eqnarray}
As illustrated in Fig.\ref{fig1}(d), phase space is 
divided into quasi-periodic and chaotic regions for $b \gg 1 $. 
The kinetic term is almost linear for $|p| < p_d$, and tends to 
the standard map for $|p| > p_d$. 
The parameters $d_1$ and $d_2$ was put 
in order to get rid of small island structures 
which may appear around the border between torus and chaotic regions. 
The smoothing factor tanh is introduced to allow analytical continuation of the map into complex plane.

The wavepacket launched at $p=0$ goes out of the torus region 
due to tunneling effects. 
Time evolution is described by the propagator  in $p$-representation $\langle p_n | U | p_0 \rangle$ 
and its leading-order semiclassical approximation takes the form as 
\begin{eqnarray}\label{semicl}
U^{sc}(p_0; p_n)  = 
\sum_{k}
A_{k}(p_0, p_n)
\exp \left\{ \frac{i}{\hbar} S_{k}(p_0, p_n)\right\}, 
\end{eqnarray}
where the summation is taken over all classical paths $k$ satisfying  
given initial and final momenta, 
$A_{k}(p_0, p_n)$ and $S_{k}(p_0, p_n)$ 
stand for the amplitude factor associated with the stability and the corresponding classical action, 
respectively. 

For given $\alpha \in {\Bbb R}$, 
we have a set for initial conditions of semiclassically contributing complex paths as 
\begin{eqnarray}\label{mset-alpha}
{\cal M}_n^{\alpha} \equiv 
\{(p_0, q_0 = \xi + i\eta ) \in {\Bbb C}^2 \ | \ p_0 = \alpha, \  p_n \in {\Bbb R} \  
\}. 
\end{eqnarray}
As shown in Figs.\ref{fig1}(a) and (b), there exist a bunch of complex trajectories that satisfy 
the boundary conditions imposed on the initial and final steps. 
The set ${\cal M}_n^{\alpha}$ is composed of strings, each of which correspond 
to the contribution $k$ in the sum (\ref{semicl}). 
In case of the present map, 
the number of strings is infinite even for a finite fixed time step, reflecting 
that the map contains transcendental functions, while only finitely many complex orbits 
appear in case of the map composed of polynomial functions. 
What was found in \cite{SI,SII} is, irrespective of the form of maps,  that 
the tunneling wavefunction is well controlled and quantitatively reproduced by 
special types of complex orbits. 
Such orbits appear as a chained structure 
in the set ${\cal M}_n^{\alpha}$, an example of which is displayed in Fig.\ref{fig1}(b). 
The orbits forming chained structures are shown in red color. 
It was also shown in \cite{SI,SII} that the number of complex orbits substantially 
controlling the tunneling process increases as a function of time. 
Further important features actually shown here is that 
all these approach the real plane exponentially. 
These are both explained by the fact that such orbits are 
attracted by unstable periodic orbits on real plane \cite{SII2}.
All the stable manifolds of unstable periodic orbits extended to complex plane in general 
reach and intersect an initial state $(p_0, q_0 = \xi + i\eta ) \in {\Bbb C}^2$, thereby the orbits 
launched at the initial state go out of regular regions by following stable manifolds and approach 
the real plane exponentially.

%

It is important to note that, after reaching quasi-real regions, the dynamics of complex orbits 
is almost governed by the real dynamics.  Figures \ref{fig1}(c) and (d) plot the distance from 
the real plane, and the projection of final manifolds onto the real plane, respectively. 
Three curves shown in Fig.\ref{fig1}(c) show the itinerary of complex orbits staring at 
three initial points marked in Fig.\ref{fig1}(a).  Note that each one is not the itinerary of 
a single orbit, but a representative of exponentially large number of orbits around each initial point, 
since these initial points are very close to each other. For example, the orbits starting at 
the red part in Fig.\ref{fig1}(b) behaves in the same way as the red curve 
in Fig.\ref{fig1}(c), 
so these are not distinguishable from each other.  

Along the final manifold presented in Fig.\ref{fig1}(d), $|{\rm Im}\,q_n|$ is very small, 
meaning that it almost follows the stretching 
and folding mechanism in real phase space.  
We notice that, within a single step, the orbits already go out from the torus region, 
and then go down to the real plane exponentially with being stretched in the unstable direction. 
In the final step ($n=5$ in this case), the stretched manifold is almost real. 
Therefore, if one focuses on the manifold contained in the chaotic regions,  
the situation is almost the same as what is taking place in real dynamics. 
In this way, complex orbits controlling chaotic tunneling bear an 
{\it amphibious character}\,: running through the torus region 
in purely imaginary space, and extending over the quasi-real chaotic region. 

Now, recall that the wave function in chaotic regions is dynamically localized \cite{Casati}.  
Although the semiclassical interpretation of dynamical localization is not still clear \cite{Kaplan}, 
it may be at least true that, if the semiclassical description works, the orbits contributing to the 
semiclassical sum (\ref{semicl}) should have appropriate correlations among them \cite{SI2}, 
otherwise, classical diffusion will be restored since the random phase in the semiclassical 
sum (\ref{semicl}) cancels the off-diagonal contributions. 
In the present situation 
it would be natural to expect that the quasi-real 
orbits discussed above are correlated in chaotic regions as well.

\begin{figure}
\begin{center}
\includegraphics[width=.80\linewidth]{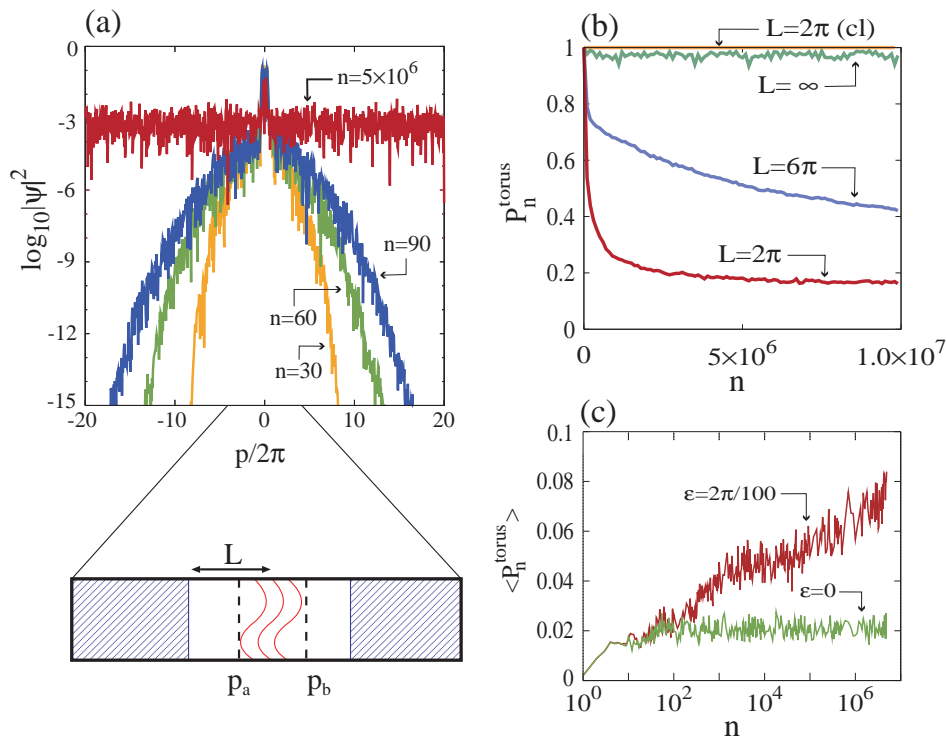}
\end{center}
\caption{{\small (a) Time evolution of wavepacket launched at $p_0 = 0$ state. 
Noise is applied only in the hatched blue regions in the schematic picture of phase space.
$a = 4+0.1\sqrt{3}$, $\omega = 2\pi (1+\sqrt{7})/10$, $d_1= -ap_d + \omega$, 
$d_2= -ap_d -\omega$, $p_d = \pi$, 
$L=2\pi, |p_a - p_b| = 1.6\pi$, $N=1000$ and $\hbar = 0.1\pi$, 
where $N$ denotes the system size. The noise intensity 
is given as $\varepsilon = 0.1\pi$. 
(b) $P^{{\rm torus}}_n$ as a function of time step $n$. 
The parameter values are the same as (a). 
Note that $P^{{\rm torus}}_{\rm eq} = 0.016$ is far below a saturated value for the case 
$L = 2\pi$.   
(c) The probability inside the torus region $P_n^{{\rm torus}}$ 
as a function of $n$ for the cases without noise ($\varepsilon=0.0$) 
and with noise ($\varepsilon = 2\pi/100$). The initial wavepacket is set as  
$\psi_0(p) = A\exp(2\pi i \eta(p))$ (for $|p| \ge 2\pi), \ =0$ (otherwise). 
Here $A$ is a normalization constant, and $\eta(p)$ = uniformly random 
in $[0,1]$. The plot is given after averaging over 10 ensembles with 
respect to the phase $\eta(p)$. 
}}
\label{fig2}
\end{figure}

To check this speculation, 
we add noise in the chaotic region and destroy the coherence predicted here. 
The experiment is done by putting the noise term to the kinetic term $T_0'(p)$ as  
\begin{eqnarray}
T'(p) = T_0'(p) + \varepsilon T_{\rm noise}'(p). 
\end{eqnarray}
Here $\varepsilon$ represents a stochastic variable obeying the Gaussian distribution
 with a variance $\sigma$. 
As schematically shown in Fig.\ref{fig2}(a), 
setting $T_{\rm noise}'(p) = 1 \ {\rm for \ } |p| \ge L, \ =0 \ {\rm for \ } |p| < L$, 
we apply noise only in the region $|p| \ge L$. 
To see the tunneling amplitude, we will monitor the probability $P^{{\rm torus}}_n$ 
defined as 
\begin{eqnarray}\label{prob}
P^{{\rm torus}}_n = \int_{p_a}^{p_b} |\psi_n(p)|^2 dp, 
\end{eqnarray}
where $\psi_n(p)$ represents the wavefunction at time step $n$, 
which is initially placed at $p_0 = 0$, and 
$p_a$ and $p_b$ are the coordinates specifying the minimum and maximal values 
of the torus region projected onto the $p$-axis (see also Fig.\ref{fig1}(d)). 
Here, we imposed the periodic boundary condition in the $p$-direction. 
As shown in Fig.\ref{fig2}(a),  wavefunction outside the torus region spreads 
as time proceeds and the profile of its tail shows that the classical diffusion process 
is recovered due to the destruction of interference which causes dynamical localization \cite{Ott,Adachi}.  
Figure \ref{fig2}(b) plots $P^{{\rm torus}}_n$, 
 clearly demonstrating a drastic enhancement 
of tunneling probability.  
Note that the ensemble of real classical orbits whose initial distribution is
set to be the same as the quantum initial distribution almost stays inside the torus 
region and does not leak out even under the same noise. 
We can see that with the increase of $L$, which makes the chaotic region 
surrounding the torus region large, the tunneling rate decreases.  
As will be reported elsewhere \cite{ITS1}, extensive computations 
reveal that the exponent obtained by fitting an exponential function to 
initial decaying interval of $P^{{\rm torus}}_n$, up to $10^{4}$
or $10^5$ in case of Fig.\ref{fig2}(b), 
decreases monotonically as a function of $L$. 
These are consistent with our interpretation 
that the presence of localized regions suppresses potentially existing 
tunneling amplitude, which is related to an exponentially large number of 
complex orbits shown in Fig.\ref{fig1}. 
At the same time, note that the noise average is not neccesarry for 
the enhancement of the tunneling: Fig. 2 (a) and (b) are such examples.

An important remark would be that 
the tunneling leakage continues but its rate $\gamma$ 
could not be fitted by a simple exponential function 
in the whole time scale. 
The rate of penetration slows down with time and  
it takes very long time, if so, to reach the equilibrium state. 
If the wavepacket spreads over phase space equally, 
the final probability should take the value $P^{{\rm torus}}_{\rm eq} = 0.016$, 
which is estimated under the assumption that the wavepacket is uniformly 
distributed over the phase space. 
(Note again that phase space in the experiment is compact since periodic 
boundary conditions are imposed on both directions.)
However, $P^{{\rm torus}}_n$ is still far above 
$P^{{\rm torus}}_{\rm eq}$, 
and the localized peak around the torus region 
remains, meaning that the wavepacket does not spread over phase space 
in an equal weight even after sufficiently long time.

%

Since an exponentially large number of complex paths exist also from 
the chaotic region to the torus region \cite{SII}, the inverse tunneling process 
should be enhanced as well. 
Indeed, as shown in Fig.\ref{fig2}(c), 
if we place the initial wavepacket in the chaotic sea, instead of the torus region. 
the tunneling flow into the torus region 
is also highly enhanced with noise being applied on chaotic seas. 
This result tells us that the chaotic region with external noise 
does not work merely as a sink or reservoir. 
If this is the case, 
the unidirectional flow from the torus to chaotic regions
is expected to take place and the tunneling back process should not be observed.

Similar strong enhancement is observed when we design the system 
so that the wavepacket moves freely in the regions $|p| \ge L$.  
The motivation to put ballistic regions is again in order to get rid of dynamical localization, 
as the case where noise is applied.
To realize it, we replace the kinetic term (\ref{kinetic}) by 
\begin{eqnarray}\label{ballistic}
\nonumber
&&T'(p) =  \frac12 a p \{ \tanh b ( p + L ) - \tanh b (p - L) \} \hspace{10mm} \\
\label{ballistic}
&+& \frac12( a p -\omega )\{\tanh b (p - p_d) - \tanh b (p + p_d) \}.  
\end{eqnarray}
The parameter $L$ controls the border from which the ballistic propagation 
begins to occur.  Furthermore, to avoid the recurrence of the wavepacket 
to the initial domain, we put the absorbing boundary at $|p| = p_{{\rm cutoff}}$.
More precisely, a projection operator $\hat{P}$, which satisfies 
$\langle p | \hat{P} |\psi \rangle = 0$ for $|p| > p_{{\rm cutoff}}$, 
and $\langle p | \hat{P} | \psi \rangle = \langle  p|\psi \rangle $ 
for $|p| < p_{{\rm cutoff}}$, is applied in each time step. 
 After penetrating through the torus region, 
the wavepacket propagates in chaotic seas up to $|p| \simeq L$. 
Then it moves ballistically in the region $|p| > L$ and is absorbed at 
$|p| = p_{{\rm cutoff}}$.
Note that the original map (2) is recovered in the limits $L \to \infty$ and 
setting $d_1 = d_2 = 0$. 

\begin{figure}
\begin{center}
\includegraphics[width=.80\linewidth]{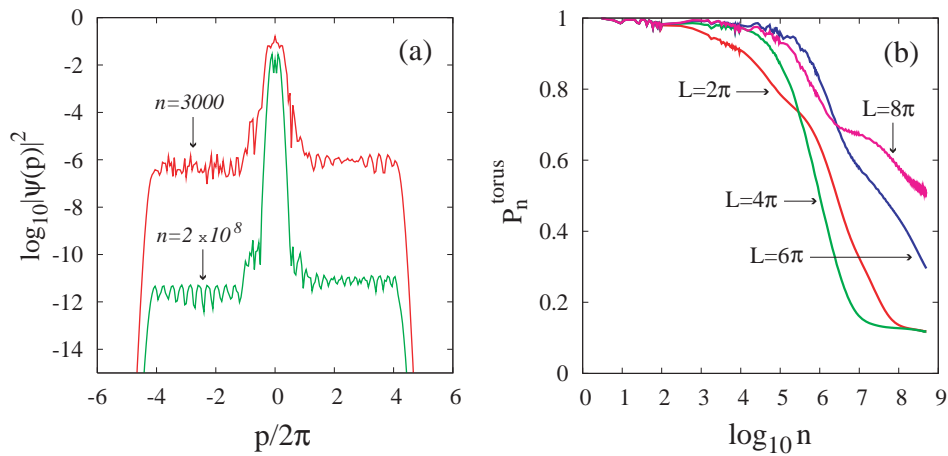}
\end{center}
\caption{{\small 
(a) The wavefunction $|\psi_n(p)|^2$ at a relatively short ($n=3000$:red) and 
a sufficiently long time step  ($n=2\times 10^{8}$:green) in case of the map 
with the kinetic term (\ref{ballistic}) with $a=4.071$, $\omega = 2.4575$, 
$N=400$, $L= 2\pi$ and $\hbar = 1/40\pi$. 
 Both are averaged over 100 steps around each time step.  
The absorbing boundary is set at $p=\pm 8\pi$, and 
the ballistic motion occurs in the blue region in the schematic phase space depicted in Fig. \ref{fig2}(a). 
(b) Long time behavior of $P^{{\rm torus}}_n$ for various $L$. 
}}
\label{fig3}
\end{figure}

We launch the wavepacket from the center of the torus region $p_0 = 0$ 
and observe the same quantity $P^{{\rm torus}}_n$ introduced above. 
As shown in Fig. \ref{fig3}, the enhancement of tunneling amplitude 
is again remarkable as compared to the system without outside ballistic 
propagation. 
We further notice that, as seen in Fig. \ref{fig3}(a), although 
the amplitude of wavefunction on the torus region is gradually reduced 
as a result of tunneling leakage, 
it is still distinguishably localized on the torus region.  
Figure \ref{fig3}(b) also exhibits that $P^{{\rm torus}}_n$ remain finite  
after exceedingly long time steps. 
Especially in cases of $L = 2\pi$ and $L= 4\pi$,  even a signature of saturation can 
be detected. 

%

\begin{figure}
\begin{center}
\includegraphics[width=.50\linewidth]{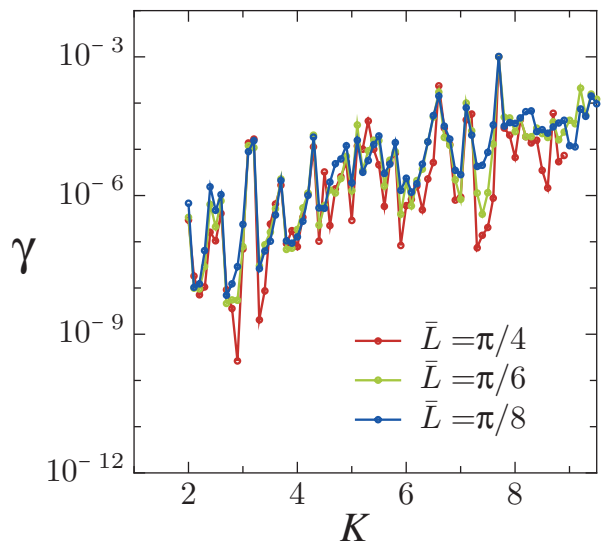}
\end{center}
\caption{{\small 
The tunneling rate $\gamma$ as a function of the kicking strength $K$. 
$\gamma$ is evaluated by fitting the initial decay of $P_n^{{\rm torus}}$
for $n < 10^{4}$ or $10^{5}$ depending on $K$. 
The length $L$ of the chaotic region without noise (see Fig.\ref{fig2}(a)) 
is scaled as $\bar{L} = L/K^2$. Here $p_d = 0.6\pi, a = 4+ 0.1\sqrt{3}$ 
and $\omega = 2\pi (1+ \sqrt{7})/10$. 
}}
\label{fig4}
\end{figure}

To see that the observed process is certainly a chaos-involved one, 
in other words, not only the localization length but also the nature of chaos, 
especially the strength of chaos controls the transition amplitude, 
we change the nonlinear parameter $K$ in our system. 
Taking into account that the localization length 
is proportional to $K^2$ \cite{CIS},  we plot in Fig.\ref{fig4}
the tunneling rate $\gamma$ as a function of $K$ 
for several scaled values $\bar{L} = L/K^2$. 
In this experiment, the initial wavepacket is again placed inside the torus region, 
The kinetic term is replaced by 
\begin{eqnarray}\label{newkinetic}
T'(p) = \frac12 a(p-p_d) + \omega + \frac12 a(p-p_d) \tanh b(p-p_d), 
\end{eqnarray}
which gives phase space whose lower part ($p < 0$) is all covered with KAM circles. 
The kinetic term (\ref{newkinetic}) is obtained by letting the second $p_d \to \infty$ 
and putting $d_1 = -a p_d + \omega$. 
Such manipulation is necessary to avoid the wavepacket penetrating 
through the KAM domains and reaching the opposite chaotic region.
The result shows that the degree of enhancement 
depends on the strength of chaos. 
Our interpretation for the result is that 
the density of stable and unstable manifolds becomes large 
with the increase of $K$, the number of 
complex paths responsible for the tunneling transition increases, 
which yields large tunneling amplitude.

%

These numerical tests, 
together with semiclassical analysis using complex 
trajectories show that 
chaotic tunneling and dynamical localization are strongly correlated. 
If the interference yielding dynamical localization is destroyed, 
potentially existing tunneling trajectories
show up, which leads to the drastic enhancement of tunneling amplitude. 
This makes a sharp contrast to the system coupled with the heat bath, in which 
quantum tunneling is suppressed \cite{Leggett}. 
We can alternatively say that torus states are sustained by surrounding dynamical 
localization, otherwise they cannot stay localized on the torus. 
Hence, amphibious states found in \cite{Ketz1,Ketz2} can be interpreted exactly 
as the flooding of chaotic tunneling trajectories. 

An important nontrivial question still not clarified is the origin of life time of 
torus states or the tunneling rate 
between regular and chaotic regions. 
It is not obvious what dynamical information is needed to specify it.  
In the semiclassical argument, 
Lyapunov exponents and the topological entropy of outside chaotic regions 
again become necessary ingredients 
since, as mentioned in the first part, dominant tunneling orbits are controlled by 
unstable periodic orbits in chaotic regions. 

It is natural to assume that 
inner torus states are more robust than 
the outer ones and the life time of the former is much longer than the latter \cite{Ketz2}. 
This would be a qualitative explanation for slowing down phenomenon, 
but 
the problem looks more subtle. 
As shown in Fig.\ref{fig1}(c), 
the orbit shown by the red curve is launched at an inner torus 
as compared to the orbits shown by green and yellow curves 
(see $\xi$ coordinate in Fig.\ref{fig1}(a) and also note that 
the initial condition is place on $p_0=0$).  
Nevertheless, it gain smaller imaginary action Im$S_n$, since it 
approaches the real plane directly, whereas the latter two orbits 
launched at outer tori take side trips, which cause additional gains of Im$S_n$. 
This suggests the inner torus has a smaller life time than the outer one and so 
the tunneling rate does not necessarily follow a simple order even in a clean setting
as the present model.

It is also necessary to examine how the life time of (complex) classical orbits inside the torus region 
is related to the tunneling rate.  The initial manifold, $\{ (p_0 = 0, \xi + i \eta) \in {\Bbb C}^2 \}$, 
represents the support of semiclassical wavefunction. It is found that the manifold 
starting at the region where the complex KAM domain dominates has a very long life time which 
has an entirely complex classical origin \cite{SII2}.

\ack
The authors thank R. Ketzmerick and A. B\"acker for their helpful comments. 
One of the authors (A.S.) is grateful to K.S. Ikeda for useful discussions.

\end{document}